\documentclass{article}

\usepackage{lineno}
\usepackage[utf8]{inputenc}
\usepackage[T1]{fontenc}
\usepackage{graphicx}
\usepackage{xcolor}
\usepackage{url}
\usepackage{booktabs}

\usepackage{algorithm}
\usepackage{algpseudocode}
\usepackage{array}
\usepackage{blindtext}
\usepackage{multicol,makecell}
\usepackage{amsfonts,amsmath,amssymb,amsthm}
\usepackage{hyperref}

\usepackage{authblk}

\title{De-Identification of French Unstructured Clinical Notes for Machine Learning Tasks}

\author[1]{Yakini Tchouka} 
\author[1]{Jean-François Couchot}
\author[2]{Maxime Coulmeau} 
\author[1]{David Laiymani}
\author[3]{Philippe Selles}
\author[3]{and Azzedine Rahmani}

\affil[1]{Femto-ST Institute, Université de Franche-Comté, CNRS, France}
\affil[2]{Université Technologique de Belfort-Montbéliard, France}
\affil[3]{Nord Franche-Comté Hospital, France}

\begin{document}



\newcommand\JFC[1]{{#1}}
\newcommand\DL[1]{{#1}}
\newcommand\TODO[1]{{#1}}

\newcommand\YT[1]{{#1}}

\theoremstyle{definition}
\newtheorem{Definition}{Definition}[section]

 \maketitle

\begin{abstract}
    Unstructured textual data are at the heart of health systems: liaison letters between doctors, operating reports, coding of procedures according to the ICD-10 standard, etc. 
The details included in these documents make it possible to get to know the patient better, to better manage him or her, to better study the pathologies, to accurately remunerate the associated medical acts\ldots All this seems to be (at least partially) within reach of today by artificial intelligence techniques.
However, for obvious reasons of privacy protection, the designers of these AIs do not have the legal right to access these documents as long as they contain identifying data. De-identifying these documents, i.e. detecting and deleting all identifying information present in them, is a legally necessary step for sharing this data between two complementary worlds. 
Over the last decade, several proposals have been made to de-identify documents, mainly in English. While the detection scores are often high, the substitution methods are often not very robust to attack. In French, very few methods are based on arbitrary detection and/or substitution rules.  
In this paper, we propose a new comprehensive de-identification method dedicated to French-language medical documents. 
Both the approach for the detection of identifying elements (based on deep learning) and their substitution (based on differential privacy) are based on the most proven existing approaches. The result is an approach that effectively protects the privacy of the patients at the heart of these medical documents. The whole approach has been evaluated on a French language medical dataset of a French public hospital and the results are very encouraging.
\end{abstract}

\section{Introduction}

Omnipresent in the fields of finance, transport, information - the list is necessarily incomplete - Artificial Intelligence (AI) governs our lives. The field of health is no exception, and even on unstructured data (e.g. textual), which is reputed to be the most difficult to manipulate. Problems that were inaccessible a short time ago are becoming soluble, such as the search for similar patient files, ICD-10 classification~\cite{amin2019mlt, neveol2018clef}, hospital readmission prediction~\cite{hasan2010hospital}, patient clustering~\cite{huang2019patient} \ldots.

However, these processes can be carried out by computer specialists in deep learning from the data science and big data professions, but not yet by doctors. It therefore appears necessary to "share" the data between medical actors and data science specialists. Because of the critical nature of the data involved, this sharing implies a de-indentifaction process which can only take place within a legal framework that governs the actors in the medical world. Institutional official rules are for instance U.S. Health Insurance Portability and Accountability Act (HIPAA)\cite{cohen2018hipaa} and the European General Data Protection Regulation (GDPR)~\cite{gdpr}.
Does a technical implementation of this de-identification that respects these rules is possible? Does it allows documents to be sufficiently rich and not too degraded to be used by A.I. algorithms for automatic ICD-10 classification for example? 

This work focuses on the ability to share textual medical documents, often written by doctors and can take the form of operating reports, clinical notes or biological examination results. 
To facilitate privacy protection, de-identification methods~\cite{levine2003identification,lafkysafe,google2020,bourdois:hal-03241384,Dernoncourt2016} have been proposed as a process to remove or mask any type of Protected Health Information (PHI) from a patient, so that it becomes difficult to link an individual to its data. The type of information that constitutes PHI is defined in part by the privacy laws of the relevant jurisdiction. For example, HIPAA regulations define 18 categories of PHI including names, geographic locations and telephone numbers. In Europe, since the GDPR does not provide such PHI definitions, most research uses the HIPAA definitions. 

A de-identification process can thus be summarized as an algorithm with two main phases.
The former is detecting all compromising information (names, addresses, ages, dates, numbers) or equivalently as a Named Entity Recognition (NER) phase. 
The latter consists in replacing these elements by simple substitute data or more complex context-specific labels, classically denoted as Named Entity Substitution (NES) phase. 
The difficulty here is that the de-identification process (both NER and NES) must be balanced between too much removing (limiting the data usefulness for downstream tasks  such as ICD10 classification or clustering) and not enough removing (allowing the public releasing of PHI information). 


\DL{The de-identification of French clinical text documents faces two main challenges. 
First, recent Deep Learning based approaches have shown real progress whatever the chosen metric (precision, recall)~{\cite{Dernoncourt2016}}. 
However, to be implemented, these approaches require dedicated datasets of sufficient size.
Unfortunately it appears that such medical datasets exist in English~{\cite{mimic2016}} but not in French. To overcome this problem, we propose an hybrid process involving a statiscal approach and a BERT-based deep learning approach. We also build our own dataset composed of a general corpus (based on Wikipedia) and  some $375$ medical notes manually annotated by us. For the NES phase, the foundation is Differential Privacy (DP) as introduced in~{\cite{dwork2006calibrating}} and particularized in Local Differential Privacy (LDP)~{\cite{duchi2013local}} when acquiring consecutive individual data.}

\DL{The second challenge of this study is the evaluation of the de-identification and more precisely how to certify that the process has remained balanced and not too destructive. For this, we decided to look at how the ICD-10 classification (10th version of the International Classification of Diseases~{\cite{icd10_wikipedia})} can be impacted by the de-identification procedure.
We collaborate with the Nord Franche-Comt\'e Hospital (HNFC), a mid-size hospital at the east of France and their coders to classify a dataset of non de-identified clinical notes and then implement an automatic association by machine learning on the original dataset and its de-identified version.}

\DL{To our knowledge, no complete de-identification technique involving BERT-based architectures and Differential Privacy exists for French clinical text documents. The experimental results we obtained are very encouraging (state of the art for the french NER phase) given the difficulty of obtaining french medical dataset, 
In summary, the contributions of this paper are as follows:
\begin{itemize}
    \item A complete de-identification tool (NER and NES) capable of labeling and substituting PHI attributes in French unstructured medical records.
    \item For Named Entity Recognition phase, we developed a new hybrid system involving statistical method and BERT-based deep learning approach to overcome the dataset lack problem.
    \item A substitution strategy based on differential privacy that combines security and utility in medical records.
\end{itemize}
}

The paper is organized as follows. Section~\ref{section:Context} details the context of our study and related work on medical document de-identification. Section~\ref{section:FrenchNER} described the NER part of our process. We detail the datasets, the models and the overall architecture we used in our approach. Experimental results are also presented. Section~\ref{section:Surrogate} focuses on the NES phase and the substitution strategies of each HIPAA are formally described.
The  evaluation of our process through the ICD-10 classification task
is presented in Section ~\ref{section:GlobalEvaluation}. We end in section~\ref{section:Conclusion} by some concluding remarks and future works.

\section{Context and Related Work}\label{section:Context}

\JFC{This section presents a general state of the art on document de-identification.
It first recalls the legislative aspect motivating this work (Section~\ref{sub:rel:legal}).
Since such an approach is composed of twofold, namely NER and NES, the most recent works concerning these two steps are recalled in Sections~\ref{sub:rel:ner} and~\ref{sub:rel:nes} respectively.
It ends by describing general de-identification approaches combining these tools in Section~\ref{sub:rel:deid}.}

\subsection{Legal Context}\label{sub:rel:legal}

Prior to any handling of medical records by a party external to the institution that disposes of them, it is necessary to ensure that the confidential health information is protected.
Quoting GDPR~\cite[(Recital 35)]{gdpr}, "personal data concerning health should include all data pertaining to the health status of a data subject which reveal information relating to the past, current or future physical or mental health status of the data subject". This regulation however allows to outsource such kind of health data, but in the restricted context of 
public health as precised in Recital 54: "the processing of special categories of personal data may be necessary for reasons of public interest in the areas of public health without consent of the data subject".

But analyzing medical documents to extract codes (as in the specific context of this work) is not a public health issue. It is a global approach to optimizing hospital resources.   
Thus, this restrictive framework authorizing the use of raw health data cannot be applied here as in many other situations.

However GDPR "does not apply to anonymous information, namely information
which does not relate to an identified or identifiable natural person or to personal data rendered anonymous in
such a manner that the data subject is not or no longer identifiable"\cite[(Recital 26)]{gdpr}.
The ability to work with health data is therefore based on the fact that the medical documents can be fully de-identified. 
State  of the art of de-Identification methods for french clinical notes are recalled in the next section.



\subsection{Named Entity Recognition Related Work}\label{sub:rel:ner}

\YT{The first NER works of the 1990s were all based on rule-based techniques which typically use regular expressions manually defined. With the emergence of machine learning, statistical approaches have emerged such as the Conditional Random Field (CRF) algorithm\cite{CRFLafferty2001}. Even if this approach provides acceptable results in terms of sensibility, the obtained accuracy is not necessarily satisfactory and this is problematic in a medical document. For example, they are unable to differentiate "Mr. Charcot" from "Charcot's disease", i.e. to take into account the context.}

\YT{The recent introduction of the Transformers~\cite{attention} and BERT~\cite{bert} models has allowed a new evolution of the field. The literature is abundant and we can cite here two comparative studies of different transformer models~\cite{NERTransformersSoA} and~\cite{TransformerComparison}. with respect to a NER task.}

\YT{All previous works focused on the English language where big labeled clinical datasets exists, i2b2~\cite{i2b2dataset} and MIMIC~\cite{mimic2016} for example. This allows us to fine-tune a specific pre-trained language model (in French FlauBERT~\cite{le2020flaubert}, CamemBERT~\cite{camembert}) or a pre-trained multi-lingual model~\cite{guarasci2022bert, xlm-r}. 
Unfortunately, there are no such datasets in other languages. To build it, it is required to have access to a substantial volume of original medical data, then annotate it manually. Recently,~\cite{silvestri2022iterative} proposed an NER system on medical documents in Italian languages with an F1-score of 96.6\%. For the same reason we will train our own model on a generalist dataset which will be less sensitive to the medical context of a clinical note. We can cite, the French WikiNER 
(see next section) dataset which is derived from Wikipedia. Nevertheless, several works exist in the general field of French NER. For example in~\cite{french-ner}, the authors present a new state-of-the-art for French Named Entity Recognition in a general purpose context (not medical). They use the recent BERT  models in their French version i.e. CamemBERT~\cite{camembert} and achieve an F1-score of 90.25\% in the task of Named Entity Recognition.
As specified by the authors, a major difficulty is to obtain French labeled datasets in order to train and evaluate their models (transfer learning). In this way, to carry out their work they had to manually label their own (relatively small) dataset.} 

\subsection{Named Entity Substitution Related Work}\label{sub:rel:nes}

\YT{The most straightforward way to substitute entities is to replace each entity value with the name of the entity itself (the LOC tag for any location for instance). It has been shown~\cite{lafkysafe} that such replacement of all the 18 HIPAA entities efficiently protects the privacy of the patients/doctors: in this article, it is shown that only 2 over 32500 health notes have been indeed re-identified.
However, this kind of coarse substitution drastically reduces the usefulness of the data. For example, it may become difficult to distinguish whether the subject of certain sentences is the patient or the doctor. More problematically, the chronology of dates is in this case very difficult to re-establish since they are completely hidden. Other more accurate substitution methods have been designed for this purpose. }

\YT{
Prior to this work, the Scrub~\cite{sweeney1996replacing} systems generated surrogate texts that match the format of the original ones. 
More precisely, for dates, Scrub associates to each detected date an approximate timestamp (the nearest month, for example). For names, a lookup table is used to ensure that the same name in the original file is always replaced by the same substitute.}

\YT{In~\cite{douglass2004computer,levine2003identification} annotated PHIs from a corpus by hand and replacement carried out semi-automatically. 
More precisely, dates are shifted by a random number of weeks and years, while preserving the day of the week.
As in Scrub, each person name is replaced by its associated one, here from a list (Boston residents).
Locations are replaced by randomly selected small towns. 
Uzuner et al.\cite{uzuner2007evaluating, Uzuner2008ADF} extend the works of Douglass et al\cite{douglass2004computer}
by  replacing strings such as identification numbers and phone numbers with randomly generated digits and letters. 
For dates they maintained internal temporal relationships by shifting all dates in a document by the same number of days
and ensuring that the surrogate dates were properly formatted.}

\YT{In the de-identification approach of Deleger et al~\cite{deleger2014preparing}, names are replaced with surrogates by randomly selecting male, female, unisex and family names from pre-compiled lists. Then, all documents are parsed to store all 455 detected places (street names, city names, state names\ldots), resulting in a corpus of places.  Dates are replaced by random dates, while respecting the format, and places by places from the above corpus, while respecting the format. This approach is clearly not satisfactory because it does not resist the membership attack:  it reveals, for example, the presence of a person from a village with a particular name in the published dataset since the name of that village is likely to appear.}

\subsection{Related Work on De-Identification of (French) Clinical Notes}\label{sub:rel:deid}

\YT{Automatic de-identification is a challenging problem whose first works date back to the late 1990s~\cite{sweeney1996replacing,GuptaDe-Id}. It is a task that has become particularly important in recent years with the evolution of Natural language processing methods. There is a need in medical research to tackle textual data. To make this data accessible while guaranteeing, several studies~\cite{sweeney1996replacing,GuptaDe-Id,ciampi2022privacy, Dernoncourt2016, grouin2015possible, Uzuner2008ADF} have been carried out on de-identification.}

Nevertheless, in the medical field and to our knowledge, there is no commercial tool or large-scale deployment. One explanation of this is the difficulty of dealing with any unstructured texts and so to guarantee the complete de-identification of all PHI. As previously stated, a too strong de-identification may lead to an information loss that can be detrimental to the analysis tasks that may follow. For example: "Charcot disease" must not be de-identified while "M. Charcot suffers from vertigo..." must be de-identified. Another example is the term "PSA" which stands for a french car company and also for a blood medical exam. These few examples underline the importance of the context and the challenging nature of the de-identification problem.

For the NER phase (and for the English language), several works have focused on the use of machine learning models such as support vector machine or decision tress~\cite{Guo2006IdentifyingPH, Uzuner2008ADF, LIU2015S47}. Recent advances in the neural approach and deep learning have led to important advances. In 2016  Lample et al. in~\cite{lample2016neural} and Dernoncourt et al. in~\cite{Dernoncourt2016} proposed the first architectures based on Neural Networks for the de-identification of medical unstructured texts. Dernoncourt used a Recurrent Neural Networks (RNN) trained on two medical datasets, namely i2b2~\cite{i2b2dataset} and MIMIC~\cite{mimic2016}.
The obtained results represent state-of-the-art with F1-scores reaching 97.85\% and 99.23\% depending on the dataset they used.
In~\cite{google2020}, the authors propose a comparison studies of deep learning systems ranging from off-the-shelf to fully customized. The authors use an hybrid system based on RNNs coupled with a CRF (similar to~\cite{Dernoncourt2016, Liu2017}). The customization levels depend on the dataset and the embedding layer they used. Unsurprisingly, their custom approaches are able to deliver the most accurate results with a F1-score ranging from 97 to 99\% on par with Denoncourt et al's results. 

The most consistent studies on de-identification of medical documents in French are mainly those carried out by C. Grouin~\cite{GROUIN2014151,grouin2015possible}. However, all of the implementations are based either on CRF or on regular expression rules. 
More recently,~\cite{bourdois:hal-03241384} focuses on the de-identification of french emergency medical records. The approach consists in two steps: first the authors use FlauBERT~\cite{le2020flaubert} (see later) to classify the notes containing data to de-identified. They then compare different approaches combining rules-based techniques and \YT{Long Short Term Memory (LSTM)} (via Flair~\cite{flair}). Note that Flair was trained on WikiNER (see next section). For the evaluation phase, they have also manually annotated a relatively small number of notes (414) where only the persons names were detected.

\section{French Named Entity Recognition}
\label{section:FrenchNER}

In this section, we describe our approach for the NER phase of the de-identifcation problem of French clinical notes.

\subsection{Named Entities}

Named Entity Recognition is the task of identifying and categorizing key information (entities) in a text. An entity can be any word or series of words. 
So one must first identify which word classes may have content that could reveal personal information. Unfortunately, due to the richness of natural language and the uniqueness of many human behaviors, there is no definitive answer to this question. Some combinations of even innocuous keywords can uniquely identify a person and can thus be seen as quasi-identifiers. An acceptable answer may be to rely on what is accepted as identifiers for a specific search domain. 

For example, all the decisions published in France on the Cour de Cassation (legal area) website have had the first and last names of individuals mentioned in the decisions removed and replaced by letters.
Additional deletions of other elements that allow the identification of persons (address, telephone number, email address...) and whose disclosure would be likely to undermine their security (or that of their entourage or the respect of their private life or that of their entourage), were also carried out before the decisions were put online. 

In the field of health,  the Health Insurance Portability and Accountability Act (HIPAA)~\cite{cohen2018hipaa} provides safe harbor guidelines that define what information that can be considered as private: Private Health Information (PHI). The HIPAA categories form an acceptable consensus~\cite{prasser2017scalable,friedlin2008software,benitez2010evaluating}, even outside their field of application, which is the USA. For the sake of completeness, Table~\ref{hipa:cat} recalls these categories.


 \begin{table}[ht]
 
\begin{scriptsize}
    \begin{tabular}{l p{10.5cm}}
    1. & Names; \\
    2. & All geographic subdivisions smaller than a state, including street address, city, county, precinct, zip code, and their equivalent geocodes; \\
    3. & All date elements (except year) for dates directly related to an individual including, birth date, admission date, discharge date, death date, etc.  date of birth, date  of admission, date of discharge, date of death; and all ages greater than 89 years and all date elements (including year) indicative of that age, except that such ages  and elements may be aggregated into a single age category of 90 years or older ;\\
    4. & Telephone numbers;\\
    5. & Fax numbers;\\
    6. & E-mail addresses;\\
    7. & Social security numbers; \\
    8. & Medical record numbers;\\
    9. & Health plan beneficiary numbers; \\
    10. & Account numbers;\\
    11. & Certificate/license numbers;\\
    12. & Vehicle identifiers and serial numbers, including license plate numbers;\\
    13. & Device identifiers and serial numbers;\\
    14. & Web universal resource locators (URLs);\\
    15. & Internet Protocol (IP) address numbers;\\
    16. & Biometric identifiers, including fingerprints and voice prints;\\
    17. & Full face photographic images and any comparable images;\\
    18. & Any other unique identifying number, feature, or code.
    \end{tabular}
\end{scriptsize}
\caption{HIPAA categories~\cite{cohen2018hipaa}}\label{hipa:cat}
\end{table}


\subsection{Non Private Training Monolingual Datasets}
\label{subsection:training_dataset}

As we will see in the remainder of this section, our approach relies on the use of transformers deep learning models such as FlauBERT. Note that, we choose to use monolingual models since several studies~\cite{TransformerComparison} show that they outperform multilingual approaches. In order to specialized FlauBERT on the NER task, we must trained it with a dedicated dataset. Unfortunately ans as stated above, there are very few french NER datasets of sufficiently large size. Among them, the WikiNER dataset~\cite{wikiner}, created by Nothman et al. 
contains manually-labelled Wikipedia articles across 9 languages, namely English, German, French, Polish,\ldots 
In French the size of the dataset consists in more than 61 000 pages and more than 3 000 000 words. The annotations are of 4 main types: LOCation, PERson, ORGanisation and MISCellaneous.
Since this dataset is based on many Wikipedia pages and is unfortunately very general, it seems natural to augment this dataset with elements specific to the field studied (here the medical field) and the country (here France). Adding specificities of the domain allows indeed to hope to treat them automatically and more precisely (cf "Charcot Disease"). 
To this goal, the QUAERO~\cite{neveol14quaero} french medical corpus has been added to the former WikiNER dataset. It is not a clinical notes dataset but a selection of MEDLINE\footnote{USA National Library of Medicine's bibliographic database: \url{https://www.nlm.nih.gov/medline/medline_overview.html}} titles and EMEA~\footnote{European Medicines Agency: \url{http://opus.lingfil.uu.se/EMEA.php}} documents that were manually annotated. The annotations are of ten types: Anatomy, Chemical and Drugs, Devices, Disorders, Geographic Areas, Living Beings, Objects, Phenomena, Physiology and Procedures. It can be noticed that the QUAERO dataset does not contain any personal information.
The main characteristics (content, quantitative attributes and number of labeled items) 
of both datasets are presented in Table~\ref{tab:datasets}.

\begin{table}[ht]
\centering
    \begin{tabular}{|l|c|c|c|c|c|}
    \toprule
    Dataset & Content & Nb of & Nb of  & Nb of  & Nb of \\
    &  & characters & words & sentences & labels \\
    \midrule
    WikiNER & Generalist & 27 926 161 & 3 054 130 & 135 276 & 415 088 \\
    QUAERO & Medical & 180 919 & 17 164 & 1839 & 2562 \\
    \bottomrule
    \end{tabular}
    \caption{\YT{Main characteristics of the two used French datasets}}
    \label{tab:datasets}
\end{table}

\subsection{System Architecture}

The approach we proposed is an hybrid one combining state of the art approaches of Transformers and CRF. Indeed, an approach based only on Transformers does not seem to be adapted because of the lack of a consistent french dataset in which one can find almost all PHI. Since some entities (email addresses, phone numbers, for example) are built from regular expressions, it seems that an approach based on rules would obtain higher prediction scores on these than another tool based on learning. 
Combining rule-based approaches with supervised or unsupervised learning approaches seems to be a necessity when the objective is to increase the accuracy of the overall approach. 
Figure~\ref{generalApproach} presents the general architecture of our proposal and is detailed in 
the following sections.

\subsubsection{Deep Learning based approach}

As stated previously, we use some models based on BERT (Bidirectional Encoder Representations from Transformers)~\cite{bert}.
It is a Transformer network composed of a suite of encoders only (N = 12 or 24 depending on the version: base with 110 millions parameters or large with 340 millions parameters).  
BERT was pre-trained on a large corpus of unlabeled text including the entire Wikipedia and Book Corpus. BERT is a bidirectional model meaning  that BERT learns information from both the left and the right side of a word’s context. Since its introduction, BERT and all its "descendants" have been widely used and proved to be very efficient. In our case the choice of an architecture based on BERT was guided by the availability of pre-trained French models i.e CamemBERT and FlauBERT described hereafter.

\subsubsection*{CamemBERT and FlauBERT}
\label{subsec_camembert}
CamemBERT~\cite{camembert} is a BERT type model developed by Facebook and the INRIA in France. It has been pre-trained on a 138Gb French corpus. More precisely it is based on the RoBERTa architecture~\cite{roberta}.

FlauBERT~\cite{le2020flaubert} is another French BERT developed by the CNRS in France. It has been pre-trained on a large heterogeneous French corpus and its performances compared to CamemBERT are very close. More generally, the results obtained with both models show that a specific French language model improves the results compared to 
similar multilingual BERT models~\cite{camembert}.

One of the main advantages of these models, and of BERT models in general, is their efficiency in case of transfer learning. The idea is to use a pre-trained model such as CamemBERT or FlauBERT and fine-tune it on a smaller and more specialized dataset. We then obtained and new specialized model. In the medical research area ClinicalBERT~\cite{clinicalbert} and BioBERT~\cite{biobert} are such models, fine-tuned on a medical corpus. Unfortunately all these models are English language models. To our knowledge, there is no French medical fine-tuned BERT model.

\subsubsection*{NERDA}

\begin{figure}[ht]
    \centering
    \includegraphics[width=\columnwidth]{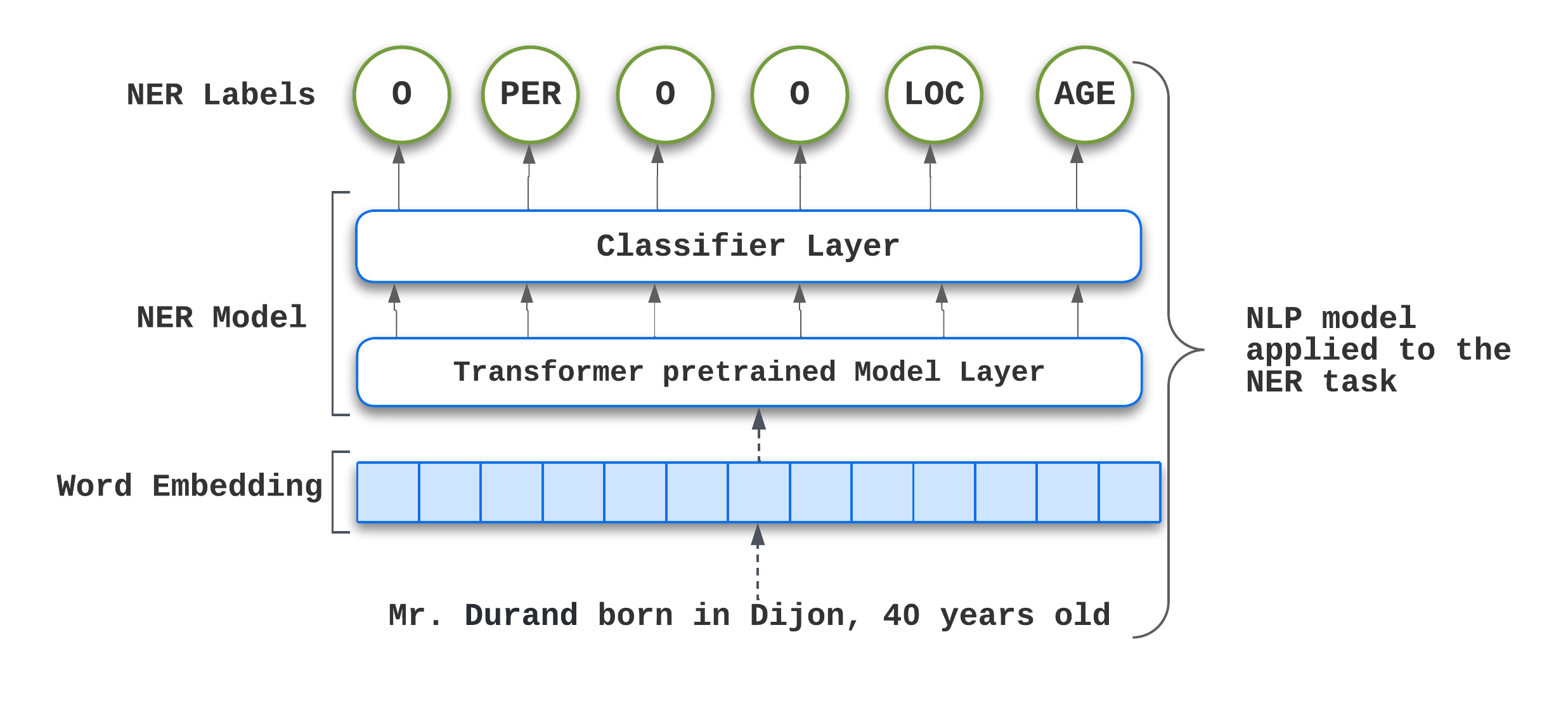}
    \caption{\YT{Deep Learning Model Architecture for NER}}
    \label{nerda:architecture}
\end{figure}

Given a fine-tuned BERT model, it is now possible to add it some layers (typically a dense layer and classification layer) to perform some NLP tasks such as text classification or Name Entity Recognition as shown in the Figure~\ref{nerda:architecture}. In this study we used the NERDA API~\cite{nerda}. 'NERDA' originally stands for 'Named Entity Recognition for DAnish'. However, this is somewhat misleading, since the functionality is no longer limited to Danish. This Python package offers an interface for fine-tuning pre-trained transformers for NER tasks. 
This architecture is presented in Figure~\ref{nerda:architecture}. In the remainder we will use NERDA and FlauBERT indistinctly.

\subsubsection{CRF based approach: MEDINA}
MEDINA is machine learning based approach using Linear chain Condition Random Field (CRF) system~\cite{grouin2013automatic} as implemented in wapiti~\cite{lavergne2010practical} for for French document de-identification.

In the context of medical notes, medical reports (written by doctors)
often do not respect any typographic rule. 
For instance, in the sentence "\textbf{jean} habite \textbf{bermont} \textbf{3 rue pierre dole}", (jean lives in bermont 3, pierre dole Street)many information can be missed by a CRF tool:
"jean" is the name of a person normally in lower case, not preceded by "Dr" or "M./Mr" and may be classified without context as a classic pant. 

\subsubsection{Hybrid System}
The hybrid approach we proposed is the following:
 
\begin{itemize}
    \item A neural approach based on BERT and more precisely one of its french version, FlauBERT for very contextual entities such as: persons, locations, organisations and medical terms. 
    \item A french CRF-based (MEDINA) to label entities such as: dates, telephone numbers, email and url. 
\end{itemize} 

\begin{figure}[ht]
    \centering
    \includegraphics[width=\columnwidth]{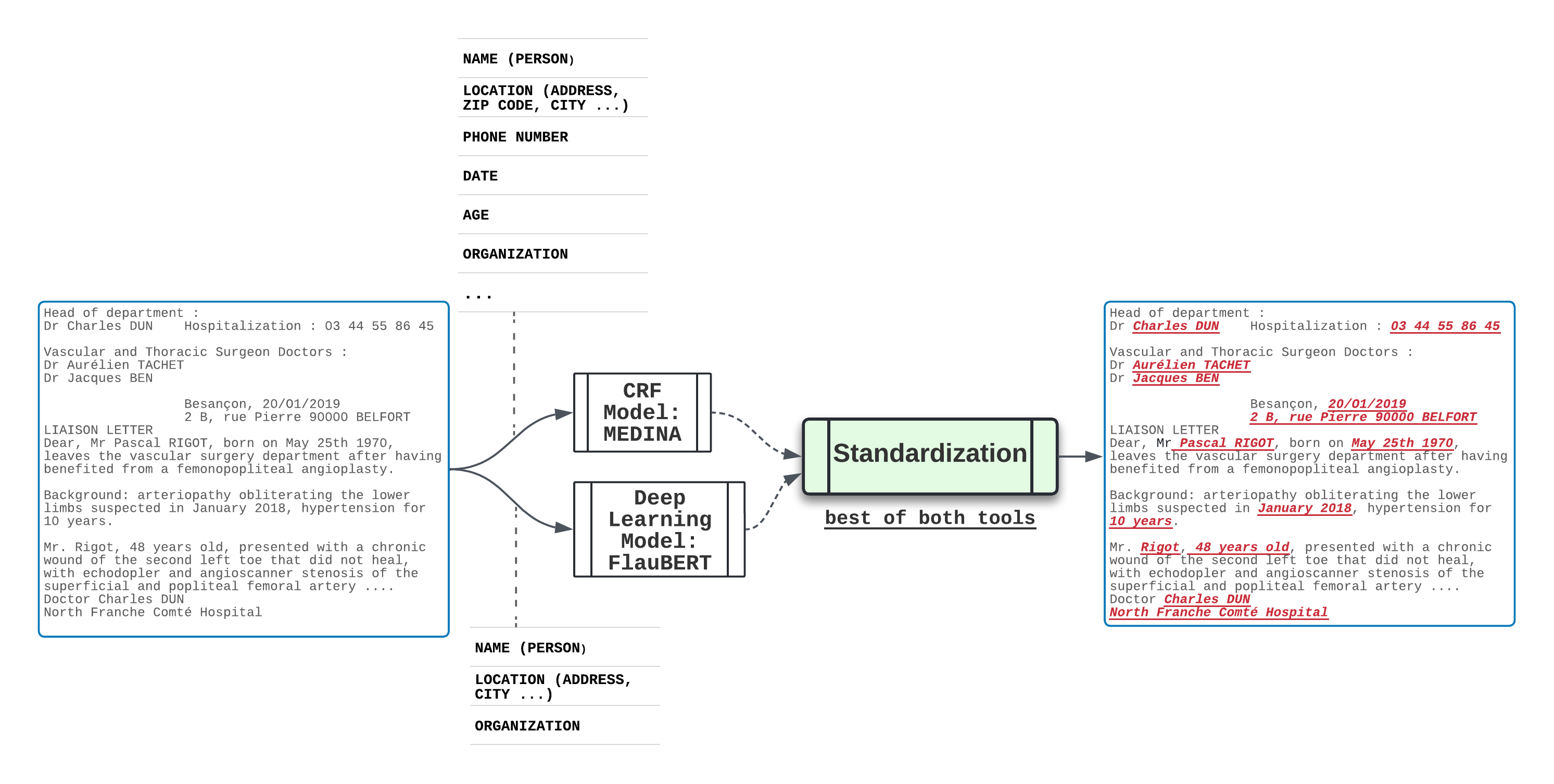}
    \caption{\YT{Our proposed hybrid approach for the NER part}}
    \label{generalApproach}
\end{figure}

Figure~\ref{generalApproach} illustrates this approach.
The input of both models is a clinical note with sensitive information.
The MEDINA model will tries to detect all attributes (person, location, age, date, phone number ...) present in the file. 
Since the deep learning model (FlauBERT) is trained only on the attributes PERson, ORGanization and LOCation (due to the WikiNER training dataset), it tries to detect only these 3attributes.
The output of each of the two models is forwarded to a decision procedure whose objective is to make the most appropriate choice according to the detected entities. 
For a given word or group of words, let $T_M$ and $T_N$ be the tags proposed by MEDINA and by NERDA respectively.  The decision procedure is based on the following rules.  
\begin{itemize}
    \item \textbf{Case 1} when $T_M=T_N$. When both tools associate the same tag, this one is simply considered as final and is returned. 

    \item \textbf{Case 2} when $T_M \neq T_N$ and $T_M$ or $T_N$ is equal to "O", the Outside value, i.e., no tag has been associated. 
    In this case, the final returned tag is the tag that is not "O". 
    This case illustrates the fact that an approach suggests a tag, which possibly identifies the patient (the ZIP code for instance), whereas the other one does not detect anything. 
    If the associated final tag was "O", no substitution would be performed later on the corresponding word. A consequence would be that we would have forgotten to clean this word and the de-identification would not be robust.
    Conversely, if the word is common, if a wording is wrongly associated with it and then replaced because of its wording, the usefulness of the de-identification is reduced, but this does not affect the patient privacy. We have favored this second scenario in the present case.

    \item \textbf{Case 3} when $"O" \neq T_M \neq T_N \neq "O"$. 
    The situation when the two approaches have associated two different 
    tags both distinct of "O"  to the same sequence occurs particularly 
    when NERDA (FlauBERT)  associates one tag in $\{\text{PER},\text{LOC},\text{ORG}\}$.
    The training dataset of this approach is indeed only based on
    this set of entities. 
    It has been shown~\cite{french-ner} that deep learning approaches are more accurate than CRF ones for very contextual entities such as persons, organizations and locations. So in this case, the final returned tag is $T_N$.
\end{itemize}

For example, in the sentence "M. \textbf{Jean} habite à \textbf{Bermont} \textbf{90400}" (Mr. Jean lives in Bermont, 90400),  MEDINA associates "PER" to Jean, "PER" to Bermont  and "LOC" to 90400 whereas NERDA associates "PER" to Jean, "LOC" to Bermont and "O" to 90400. The final association returned by the decision procedure will be Jean is PERson (case 1), Bermont is  LOCation (case 3) and 90400 to LOCation (case 2).

\subsection{Evaluating the NER approach}

\begin{figure}[ht]
    \centering
    \includegraphics[width=\columnwidth]{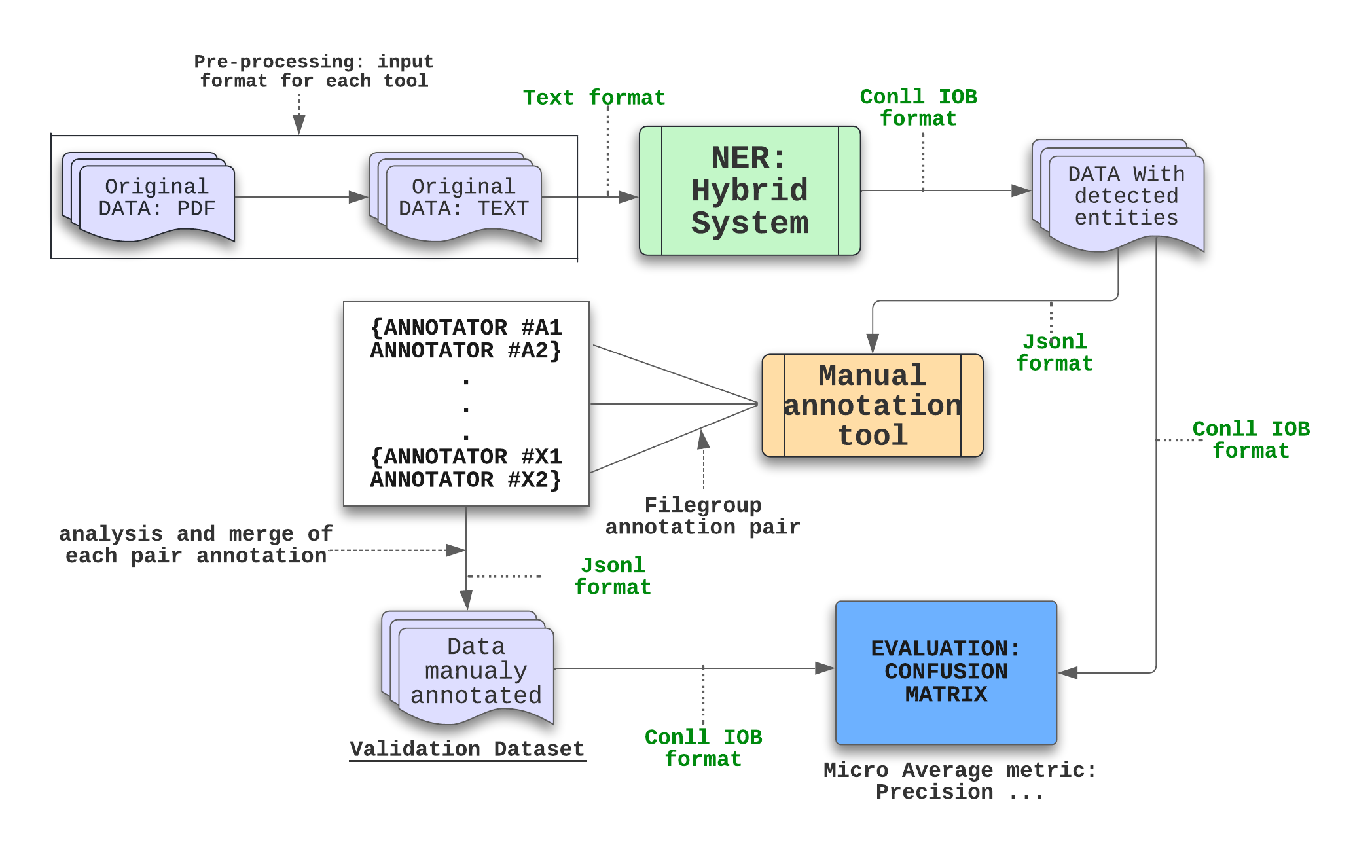}
    \caption{HNFC data set creation and evaluation process}
    \label{fig:hnfcdatasetandevaluation}
\end{figure}

This section describes the evaluation methodology we carried out for the NER phase. Figure~\ref{fig:hnfcdatasetandevaluation} summarizes the whole
evaluation approach.

\subsubsection{HNFC dataset}\label{subsub:hnfcdataset}

Evaluating our hybrid approach requires an annotated set of french clinical notes. The only way to get a such dataset (especially in French) is to manually annotate some existing notes. We then use 375 files of deceased patients from the Nord Franche-Comté Hospital. These files were pre-annotated by the hybrid system previously presented and then were manually annotated by the hospital staff with Doccanno~\cite{doccano} manual annotation tool. 
Given the automatically pre-annotated files, the annotator is responsible for checking, correcting and completing the possible errors produced by the model. HIPAA attributes were formalized to avoid ambiguity and multiple annotation criteria. For example, "Ehpad de Bermont" (Retirement Home in Bermont) can be annotated in several ways: 
Ehpad as O, Bermont as LOCation or 
Ehpad de Bermont as ORGanisation. Similarely,
"dans 3 jours" (in the 3 days) is a date while "3 x par jour" (3 times per day) is not a date. 
To minimize the risks, two annotators worked in parallel on the same files and each  annotation  pair  is  then  manually  analyzed  and  merged  into  an unique one. 
This work was performed by 6 people during 6 hours. The approximations encountered around the attributes, like the examples given above, are the result of this experimentation. 
At the end of this process, we have obtained a reference dataset with which we can use to evaluate our model. 
In the following, this dataset is referred as HNFC dataset. 
Note that, for all our tests (annotation and evaluation) we worked on site and no note came out of the hospital.

\begin{table}[ht]
    \centering
    \begin{tabular}{|l|c|c|c|c|c|c}
        \hline
        Dataset & Content & Nb of & Nb of  & Nb of  & Nb of \\
        &  & characters & words & sentences & labels \\
        \hline
        HNFC & Medical & 775 035 & 156 423 & 9993 & 23829 \\
        \hline
    \end{tabular}
    \caption{\YT{Validation dataset characteristics}}
    \label{tab:valdata}
\end{table}

\subsubsection{Comparison of NER approaches on HNFC dataset}

For evaluation, we use the classical metrics: precision, recall and F1-score. Let $TP$ be the number of true positive annotations, $FP$ the number of false positive annotations, and $FN$ the number of false negative annotations. Then, the recall $R$ is given by 
$ R = TP/(TP+FN)$, and the precision $p$ is given by  
$P = TP/(TP+FP)$. Recall and precision answer two questions about a named entity recognition tool, respectively "did we find everything we were looking for?" and "did we only label what we were looking for?". 
The F1-score metric combines precision and recall, usually by taking the harmonic mean of the two. To get a sense of the overall performance of the system, we use the micro-average of precision, recall, and F1-score. To compute the micro-average, a confusion matrix is created for all categories, and then precision and recall are computed from this table, giving the same weight to each PHI instance regardless of its category. 

To establish some baselines for our evaluations, we recall that our model has been trained on the French WikiNER dataset. The french NER pipeline from the Python Spacy library~\cite{spacy2} has also been trained on WikiNER (to which has been added the small dataset Sequoia). The model used is based on a Convolutional Neural Network (CNN) which provides a F1-score of $0.84$ for predicting 4 labels: Person, Location, Organisation and Miscellaneous.

J-B. Polle \footnote[1]{\href{https://huggingface.co/Jean-Baptiste/camembert-ner}{Jean-Baptiste Polle (2020). CamemBERT model for Named Entity Recognition task}} proposes CamemBERT-ner, a pre-trained model for french NER based on CamemBERT and trained with WikiNER. This model exhibits a F1-score of $0.89$ but the evaluation was not performed by the author on a sub-set of WikiNER but on a personal crafted dataset composed of emails and chats. 

\begin{table}[ht]
\centering
    \begin{scriptsize}
    \setlength{\tabcolsep}{1pt}
    \begin{tabular}{|l |c c c | c c c | c  c c| c c c| c c c|}
    \toprule
    Labels & \multicolumn{3}{c}{Spacy} & 
    \multicolumn{3}{c|}{CamemBERT} & \multicolumn{3}{c|}{MEDINA} &    \multicolumn{3}{c}{NERDA} &    \multicolumn{3}{|c}{PROPOSAL}\\
     &  &  & &  & NER & & & &  & & & &\\
     & P & R & F$_1$&  P & R & F$_1$& P & R& F$_1$& P & R & F$_1$ & P & R & F$_1$\\
     \midrule
    PERson & 59 & 76.8 & 67  & 89 & 99 & 93.8 & \textbf{98.2}& 97.7 & \textbf{98.2} &91.8 & 97.6 &94.6 &96.3 & \textbf{99.8} & 98\\
    ORGanisat. & 2.2 & 10.9 & 3.6  & 7. & 21.8 & 11.1 & 32.6& 24.8& 28.1 & 16.9 & 34.1 & 22.6  & \textbf{41.1} & \textbf{57.3} & \textbf{47.8}\\
    
    LOCation & 40.3 & 11.9 & 18.4  & 46 & 67.2 & 54.6 & \textbf{98.8} &81.1 & 89.1 & {75.7} & {66.3} & {70.7} & 88.4 & \textbf{95.8} & \textbf{92}\\
    \midrule
    Date & & & & & &  & 97.7 & 86.6 & 91.9 & & & & \textbf{97.7} & \textbf{86.7}& \textbf{91.9}\\
    Age & & & & & &  & 91.5 & 66.9 & 77.3 & & & & \textbf{91.5} & \textbf{66.9} & \textbf{77.3}\\
    Phone N. & & & & & & & 99.5 & 97.9 & 98.7& & & & \textbf{99.5} & \textbf{97.9} & \textbf{98.7}\\
    \midrule
    Micro Av. & 54.9 & 54.8 & 54.9 & 70.8 & 51.5 & 59.6 & \textbf{98.2} & 91.2 & 94.5 & 85.8 & 86.7& 86.3& 94.6 &\textbf{94.9} & \textbf{94.7} \\
    \bottomrule
    \end{tabular}
    \end{scriptsize}
    \caption{NER results on the HNFC dataset} \label{tab:exp:results}
\end{table}

Table~\ref{tab:exp:results} summarizes results of the different NER approaches applied to the HNFC dataset. 

For all appraoches PERson, ORGanisation and LOCation are searched in documents, since the training step of each of them is 
based on this set of tags.
MEDINA, which is a CRF model, is able to perform research for Date, Age and Phone Number in addition to.
For each tag, the higher results for Precision, Recall and F1-score are emphasized in bold. 

Let us first focus on the common tags.
For all recalls, the highest recall score is the one provided by our hybrid proposal. Notice that MEDINA provides more precise results in PERson and LOCation categories 

As far as the detection of dates, ages and telephone numbers is concerned, it is stated that these tags are absent from all learning-based approaches. Only MEDINA detects them (partially). Due to rule 2 of the decision procedure of our approach, the entities detected in this set by MEDINA will systematically be returned as is. The results are therefore those of MEDINA.  

The last line Micro Average summarises the contribution of this hybrid entity detection approach: our proposal is globally the one that detects entities most often (highest recall), without neglecting precision (with the highest F1-score).
Notice that the aggregated micro average recall score is 94.9\%. 
The drawbacks to this picture are clearly the detection of temporal aspects (dates, ages) and of organisations.
From a privacy point of view, the incomplete identification of dates is problematic as these temporal elements are quasi-identifying (see section~\ref{sub:dates}).
As far as organisations are concerned, the hospital partner thinks that this data is less sensitive.

As detailed in the architecture section our model is a hybrid system that combines a machine learning method (MEDINA) and a deep learning model (FlauBERT or NERDA). MEDINA as illustrated in the Figure is in general less accurate in recall than in precision i.e. the model misses the attributes that need to be detected more than it detects the words that are not attributes (Location $\Rightarrow$ 0.811,  Organization $\Rightarrow$ 0.248 ). It is very important to have a better recall in a privacy context. it is better to detect all the confidential information that needs to be detected than to detect the ones that are not. Our deep learning based system (NERDA) allows us to improve this. It outperforms MEDINA in recall, which improves the results in these hybrid model cases. 
The improvement is due to our hybrid system (decision procedure) which allows us to correct the errors of one of the methods by the other. The aggregation of the associations increases our accuracy of detection on the one hand and of association on the other hand. 

Let's take for instance
the sentence 
"Mr. \textbf{Jean} lives in  \textbf{Bermont}, \textbf{90400}". MEDINA associates Jean to PERson, 
Bermont to PERson and 
90400 to LOCation 
whereas 
NERDA associates 
Jean to PERson, 
Bermont to LOCation and 
90400 to Outside. 
Due to the decision procedure, 
our hybrid system will finally associate  Jean to "PER", Bermont to "LOC" and 90400 to "LOC". 
It will go from 66\% of precision for each tool to 100\%. This hybrid system not only improves the results but also allows to take into account all the main attributes (HIPAA) present in the medical documents.

\section{Surrogate Generation Strategies}
\label{section:Surrogate}

This section describes the Named Entity Substitution  step of our approach 
for the de-identification problem of French clinical notes. 
It starts with related works of the field and continue by showing that PHI can be divided into two types of categories (Section~\ref{sub:nes:cat}). Random based substitution are detailed in Section~\ref{sub:nes:random} whereas Local Differential Privacy based 
approaches are defined in Section~\ref{sub:nes:ldp}. 

One of the systems used in recent research is the system developed by Stubbs et al.\cite{stubbs2015automated}. The authors use, for names, numbers and letters the system described by Deleger et al.\cite{deleger2014preparing} For geographic locations, they use a pre-compiled list of different types of geographic locations, and a random choice to generate the surrogates. For the dates a uniform date shift with a random number of days was applied. This is the system used in medical corpora available today for research such as the 2014 i2b2/UTHealth\cite{kumar2015creation} corpus. The French document de-identification tool (MEDINA\cite{grouin2013automatic}) also uses the recommendations of Stubbs system. This paper uses a method that is based for some categories on the Stubbs system and makes contributions in the generation of dates, ages and geographical locations.

\subsection{Splitting Categories}\label{sub:nes:cat}

For some categories, generating consistent surrogates is straightforward.  
For example let us first consider phone numbers, URLs, email addresses, are alphanumeric strings of numbers, letters and special characters.  
All these elements are of course strongly identifying, but are not clearly linked to health data.  
Random substitutions can be applied on all these entities to ensure privacy without any consequence on utility provided the text format is respected.

Other categories are more problematic.
Dates and ages, which are temporal data, clearly explain the chronology of medical developments.
Locations can affect pathologies: for instance, some cities have high radon levels which can significantly increase the risk of cancer.
Randomly substituting these data does not make sense as they directly affect health.  

The approaches developed in the literature concerning dates can be summarized as generalizations of a shift or addition of a bounded noise. 
Both cases are considered as non robust~\cite{machanavajjhala2007diversity, dinur2003revealing}. Introduced by Dwork~\cite{dwork2006calibrating}, differential confidentiality is a mathematical context that allows the publication of an individual's information while respecting the latter's privacy. It therefore guarantees the confidentiality of the individual during the process of disseminating information by means of queries on a database containing or not containing the data of the latter. 
In our case, we want to clean up all the dates of a document in order to be able to use it as many times as we want.  This amounts to considering that the patient  has a mechanism that he applies locally to his documents. This is known as local differential confidentiality (LDP)~\cite{duchi2013local}, the definition of which is given below

\begin{figure}[ht]
    \centering
    \includegraphics[width=\columnwidth]{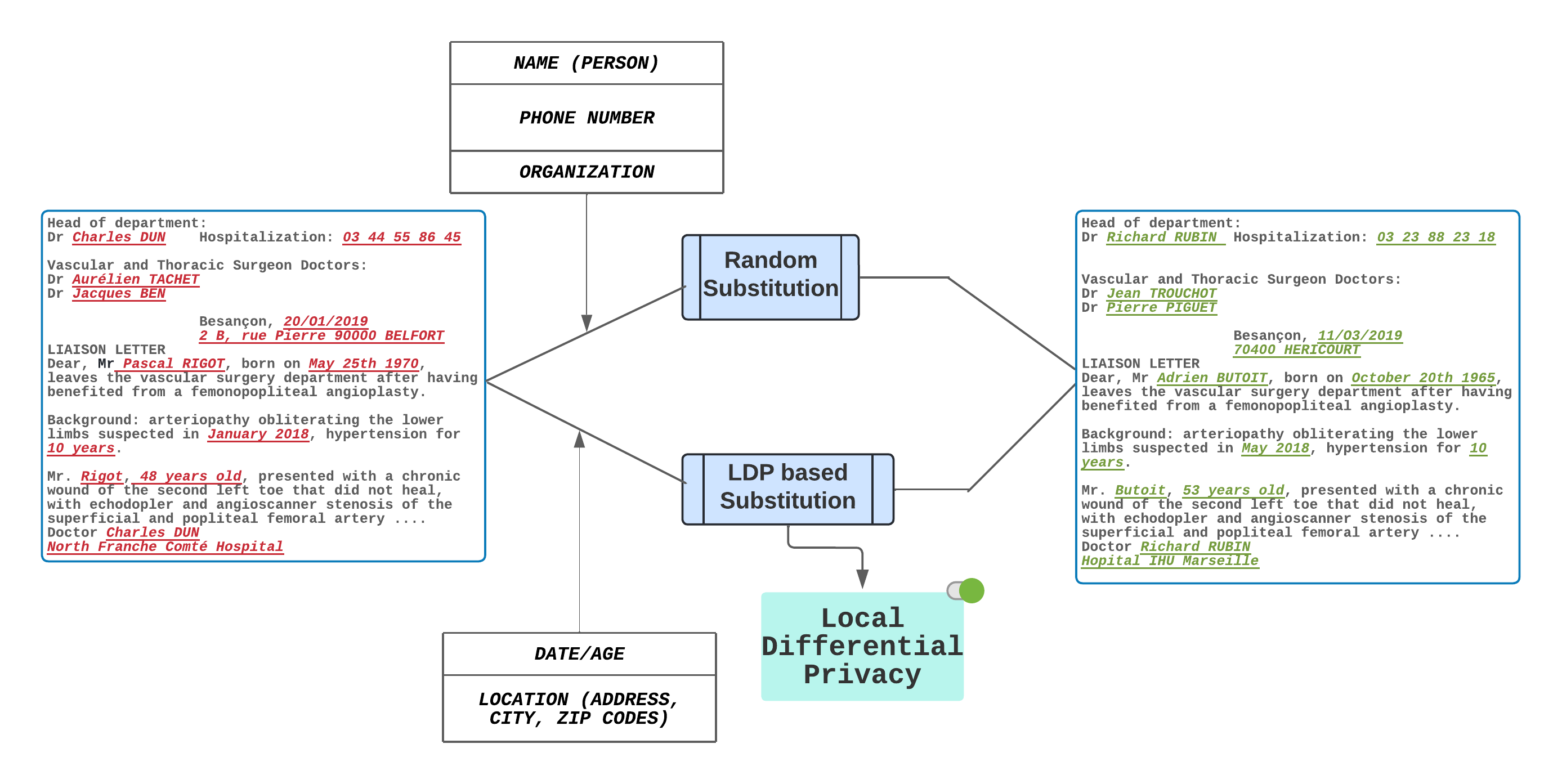}
    \caption{Substitution process}
    \label{label:substitution}
\end{figure}

The next two sections detail these two faces of sanitization and are summarized in 
Figure~\ref{label:substitution}.

\subsection{Random Substitutions}\label{sub:nes:random}

Among categories detected by the NER step, most of them contains values which  can be replaced by random values without affecting further processing. For example, generating substitutes for categories such as emails, URLs, phone numbers ... corresponds to random substitution: a phone number can be replaced by any random phone number.  

\begin{figure}[ht]
    \centering
    \includegraphics[scale=0.7]{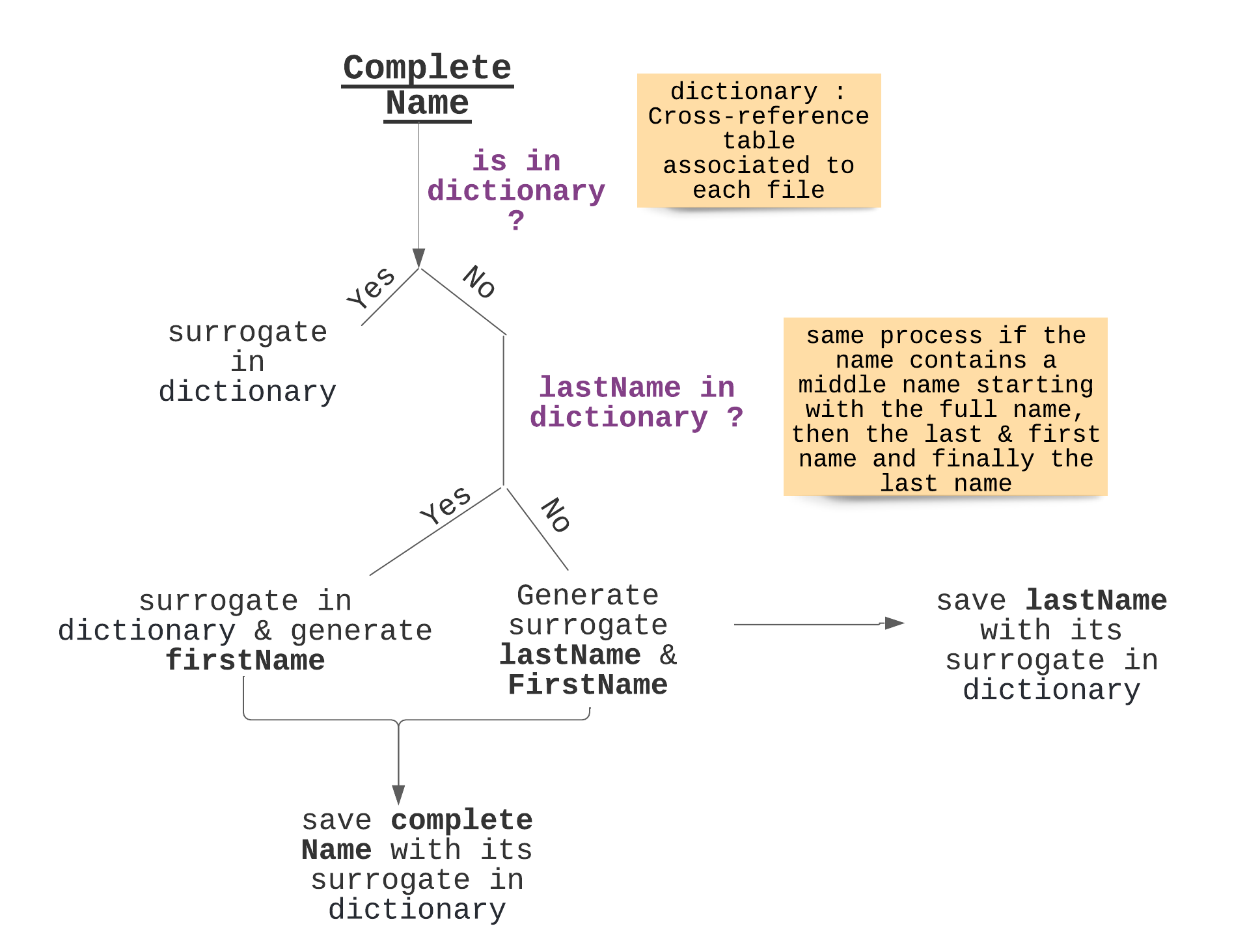}
    \caption{Name Surrogate Strategy}
    \label{name_surrogate}
\end{figure}

The names are a little different because it is interesting to preserve the affiliation within the documents.
The algorithm is illustrated in the Figure~\ref{name_surrogate}.
First, a lookup table is created for each file. 

We start by checking if the current full name is in the dictionary. If it is not, we check if the corresponding surname is in the dictionary, if it is not, it means that we have not processed it yet. Its processing consists in generating its substitute (last name \& first name) and in registering it in the dictionary. Moreover, we only register the surname that corresponds with its substitute. If on the other hand its surname is in the dictionary, we recover its substitute and we generate only the first name and we record the couple (last name \& first name) in the dictionary. Finally, if the full name is in the dictionary, we simply retrieve its substitute.

\subsection{Local Differential Privacy Context} \label{sub:nes:ldp}

Initially formalized in~\cite{duchi2013local},  local differential privacy (LDP) 
ensures individual's privacy during the data collection process.  
A formal definition of LDP is given in the following:

\begin{Definition}[$\epsilon$-Local Differential Privacy]\label{def:ldp} A randomized algorithm ${\mathcal {A}}$ satisfies $\epsilon$-LDP if, for any pair of input values $v_1$ and 
$v_2 \in \text{Domain}(\mathcal{A})$ and any possible output $y$ of ${\mathcal {A}}$:

\begin{equation*}
    \Pr[{\mathcal {A}}(v_1) = y]\leq e^{\epsilon }\cdot \Pr[{\mathcal {A}}(v_2) = y] \textrm{.}
\label{eq:ldp}
\end{equation*}
\end{Definition}
This property is a strengthening of the original centralized model of differential privacy (DP)~\cite{dwork2006calibrating} since it applies on each records whereas the original one applies on each query on the whole dataset.
Roughly speaking, a small-scale noise should suffice for a weak privacy constraint (corresponding to a large value of $\epsilon$ ), while a greater level of noise would provide a greater degree of uncertainty in what was the original input (corresponding to a small value of $\epsilon$ ).

Many mechanisms verify this local differential confidentiality.
They can be classified according to the type of considered data (categorical, real, integer), and according to their usefulness with respect to a question. 
We recall below the Laplacian mechanism (introduced by Dwork~\cite{dwork2006calibrating}) because of the simplicity of its implementation on real data.

\begin{Definition}[Laplacian mechanism in an interval of amplitude $\Delta$]
In the Laplacian mechanism, a numerical value $v$ is transformed into a numerical value ${\mathcal {M}}_{\mathrm {Lap} }(v,\Delta,\epsilon)$
 such that 
\begin{equation}
    {\mathcal {M}}_{\mathrm {Lap} }(v,\Delta,\epsilon )=v+\mathrm {Lap} \left(\frac{\Delta}{\epsilon }\right)
\label{eq:lapl}\end{equation}

\noindent where $\mathrm {Lap} \left(\frac{\Delta}{\epsilon }\right)$ is the Laplace distribution centred in 0 and whose scale parameter is $\frac{\Delta}{\epsilon }$.
\end{Definition}

The addition of noise in the Laplace mechanism ensures local differential $\epsilon$-confidentiality. 
As defined, this noise depends on the amplitude of the input data interval and of $\epsilon$.

$\epsilon$-LDP provides several important properties, e.g., immunity to post-processing  (and thus robust against any supplementary knowledge based attack) and composition~\cite{dwork2014algorithmic}. 
That is, if mechanisms are sequentially applied  to many elements inside a document of a person
the whole budget $\epsilon$ is  equal to the sum of all the budgets of all mechanisms.

Generally, data sanitization mechanisms verifying $\epsilon$-LDP are optimised according to the nature of the data taken as input. For example, a mechanism adding noise according to a Laplace distribution to numerical data can be applied to integers, but its utility will be reduced compared to the exponential mechanism dedicated to discrete data. 
Thus, for each of the categories to be sanitized, a careful selection of the mechanism to be used must be made.
In the context of medical notes de-identification, there remains 2 categories namely location and date elements, on which a $\epsilon$-LDP sanitization process will be applied on. Location is a spatial data whereas age/date may be seen as a number. Since the domains of both are not equal, a 
mechanism dedicated to each of those is thus developed.
For any of them, the privacy budget $\epsilon$ should be processed with the maximum care.


Defining the $\epsilon$ leakage budget for each algorithm can be achieved by sharing the $\epsilon$ budget among all of them. For instance, in our medical context, starting from the global $\epsilon$ budget that is allocated to whole sanitization process, $\frac{\epsilon}{4}$ can be associated to location, and $\frac{3\epsilon}{4}$ to date\&age,
but any other partition of the $\epsilon$ original budget would provide the same level of protection.
Note, however, that some attributes do not have the same degree of criticity or the same discriminating ability. For example, in a set of textual documents where the whole cohort concerns retired people, age, which is in a rather low amplitude class, is not as critical nor as discriminating as in a set of adult people.

When the attributes have a similar level of criticality and therefore the choice of the partition is more open, it should be made with respect to the accuracy that can be obtained through predictions once the dataset is sanitized. 
However, these predictions can only be made once the dataset has been sanitized.
The question arises of finding optimal parameters w.r.t prediction accuracy knowing that any re-reading of the personal data will itself reduce the leakage budget.   

Notice finally that other combination are possible for instance share the $\epsilon$ budget proportionally  to the number of detected occurrences of each category, proportionally of the number of entities concerned (Date, Age, Location \ldots).


\subsection{Sanitizing locations}
To solve the problem of privacy location, 
we used the concept of Geo-Indistingui\-shability~\cite{andres2013geo, chatzikokolakis2013broadening} which is based on $\epsilon$-LDP.  
It consists briefly in retrieving, for a given location, a location at a certain distance from it by differential privacy. This process is detailed in the algorithm~\ref{alg4:cap}.
The association $(Z,Y)$ between the original location $Z$ and the sanitized one $Y$ is stored and used in the whole document if $Z$ appears at least twice.    
This step, often denoted as memoïzation, allows to resists to a possible averaging attack. 

\begin{algorithm}
\caption{Sanitizing locations}{\label{alg4:cap}}
\scriptsize
\begin{algorithmic}
\item
\begin{enumerate}
    \item $F$ is a local list of cities $City(long, lat)$ in the local area with there longitude and latitude data.  
    \item Given a location $Z$ to sanitize. Extract its $(long,lat)$ data from  locations $F$ 
    \item From $Z$, generate of $Z'(long,lat)$ by applying the \textbf{Geo-Indistinguishability}\cite{andres2013geo,chatzikokolakis2013broadening} algorithm 
    \item Let $Y$ be the location in $F$ closest to $Z'$.
    \item Save the mapping $(Z,Y)$ in a correspondence table for this document.
\end{enumerate}
\end{algorithmic}
\end{algorithm}

\subsection{Sanitizing dates and ages}\label{sub:dates}

The objective with dates is to preserve the temporality of events in the medical document to gain information during a second analysis while respecting the privacy of patients. 
In the public medical data sets available for research (i2b2~\cite{i2b2dataset} and MIMIC~\cite{mimic2016}) a uniform shift of a randomly drawn number of days is performed on the dates. 
This process poses confidentiality problems. 
To evaluate this approach 
on the HNFC dataset (see Section~\ref{subsub:hnfcdataset}) 
and for each document, all temporal elements (dates, ages converted to dates) are 
stored as an ordered sequence 
$S_e =[e_0, e_1 ,e_2, \dots,e_n]$, from to the current date $e_0$, the most recent in the document $e_1$ and the oldest one $e_n$.
A second sequence $S_i$ of intervals is generated 
$S_i =[e_0-e_1, e_1-e_2, \dots,e_{n-1}-e_n]$, by computing the differences  (expressed in days) between two consecutive temporal elements of $S_e$.
Finally let $S'_i$ be a copy of $S_i$, but the first element which is removed. 
Notice that applying a uniform shift (as done in~\cite{uzuner2007evaluating, Uzuner2008ADF}) on all the temporal dates will not modify the sequence $S'_i$.
Note that, in the HNFC dataset we used for our evaluation, all the 375 documents provide distinct $S_e$ and  
only 8 out of 375 documents did not contain unique sequences of intervals $S'_i$.
We conclude that approximately 98\% of the chronological intervals present in documents are unique and therefore very strongly identifiable.

As detailed above, for each document, sequences $S_e$ and $S_i$ are computed.
Each interval in $S_i$ is a number of days, \textit{i.e.} a numerical value.
Local differential privacy mechanism, like Laplacian one,
can thus been applied on it.
More precisely, the bounded Laplace mechanism~\cite{holohan2018bounded} is applied here to avoid 
negative noise while preserving privacy. 
Concerning the budget for each interval, several questions arise: 
should the global budget allocated to the Date category be distributed uniformly? What are the most compromising dates? We see that the older the date, the more sensitive it is (Date of birth for example). 
How to deal with a huge interval amplitude $\Delta$ (100 years) 
in this context?

\begin{figure}[ht]
    \centering
    \includegraphics[width=\columnwidth]{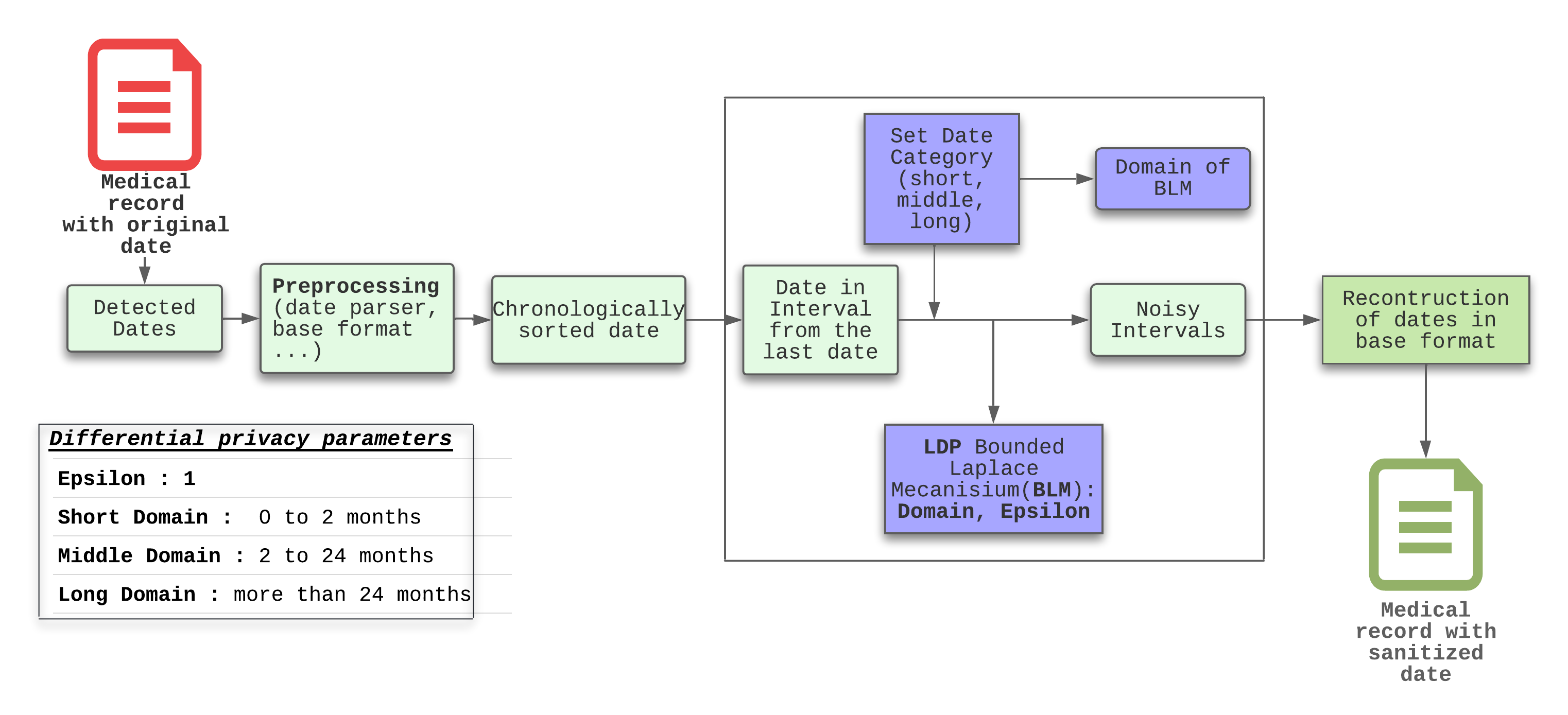}
    \caption{Date Surrogate Strategy}
    \label{date_surrogate}
\end{figure}

To solve this problem, dates have been classified into categories (Short, Medium, Long term) and for each category, an interval amplitude $\Delta$ is computed.
More precisely, $\Delta_S=61$, $\Delta_M=660$ and $\Delta_L=36,000$.
Each temporal interval is associated to one of these 3 categories. In case of ambiguity (a date in the short term and a date in the medium one, for example), the smallest category is associated to each interval.  
The global budget $\epsilon$ is split uniformly between  all the dates. 
Then, a Bounded Laplacian Mechanism~\cite{holohan2018bounded} with parameter $\epsilon_i$ and $\Delta_i$ is applied to each interval $i$, where $\Delta_i$ an
interval amplitude associated to $i$.
This approach is detailed in the algorithm~\ref{alg3:cap} and illustrated in Figure~\ref{date_surrogate}.


\begin{algorithm}
\caption{Sanitizing dates and ages}{\label{alg3:cap}}
\scriptsize
\begin{algorithmic}
\item
\begin{enumerate}
    \item Identification : Identify all the temporal elements $e$ (dates, ages)
    of the document
    \item Normalization : Normalize each $e$ in a standard format (ex. dd/mm/yyyy)
    \item Establish the chronology of the latter (classify from the first date to the last including the current date), \textit{i.e.}, compute $S_e$
    \item Define Date category (short, medium, long term)
    \item Compute the interval sequence $S_i$  between consecutive dates in $S_e$
    \item Apply to the intervals the local differential privacy with a Bounded Laplacian noise where $\Delta$ is the category amplitude
    \item Reconstitute dates from the current date
\end{enumerate}
\end{algorithmic}
\end{algorithm}

\section{De-Identification Utility Evaluation Regarding to a ML Task}
\label{section:GlobalEvaluation}

\begin{figure}[ht]
    \centering
    \includegraphics[scale=0.6]{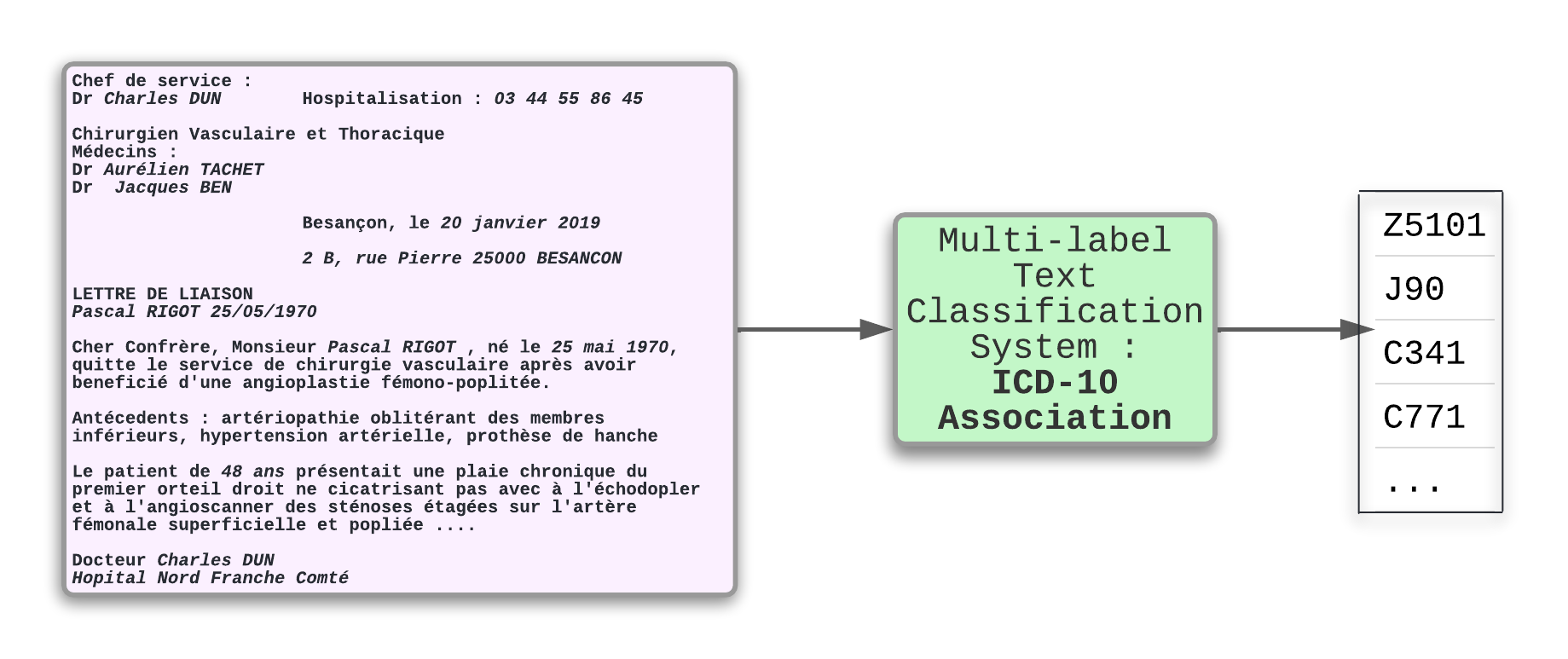}
    \caption{ICD-10 Code association}
    \label{icd10}
\end{figure}

\YT{To evaluate the medical usefulness of the documents de-identified by the method proposed in this paper, 
the ICD-10 code association task has been chosen. 
This work is often performed by ICD coding specialists in health centers and consists of manually going through medical documents and associating a set of codes to them as illustrated in Figure \ref{icd10}.} 

{Used dataset contains medical reports representing the stays of deceased patients
in the Nord Franche-Comt\'e Hospital (HNFC). 
For each stay is associated the set of manually corresponding ICD-10 codes.
To give an order of magnitude on this data set, there were 1336 documents, representing 1336 stays, 1272 different codes, with a average of 9.2 codes per stay and a standard deviation of 4.    
However, some codes are extremely rare and thus hard to associate. Thus to overcome the imbalance of our dataset (1336 inputs compared to 1272 codes) and to make this task more affordable, the codes were grouped into ICD-10 chapters\footnote{\url{https://icd.who.int/browse10/2019/en}}. 
Starting with the original dataset, we first obtain a dataset, further denoted to as ORIG-HNFC, containing stays and there 
associated chapters. 
Among the 22 original chapters, 20 are present in the ORIG-HNFC dataset.
A second dataset is composed of de-identified documents following methods described in this article. 
and there associated chapters too, which are the same ones than the ORIG-HNFC dataset. 
This dataset is further denoted to as DE-ID-HNFC.}

\JFC{The objective is to exhaustively run up-to-date Machine Learning ICD-10 association approaches on both dataset and to compare the utility loss due to de-identification. 
In the last few years, several studies have been conducted to address Machine Learning ICD-10 association, but only one in French-language, 
namely~\cite{dalloux2020supervised}. It is a supervised learning method for multi-label text classification with Convolutional Neural Networks. FastText vectors~\cite{fasttext} were used by the authors to encode the documents. 
We re-implemented, trained, and evaluated this model on the two datasets previously mentioned 
with the same parameters and hyperparameters. 
The performance measures (Precision, Recall, and F1-score) of the evaluation for each dataset are presented in Table \ref{result}. 
The micro-average system is used to obtain the aggregation of the performances.}

\begin{table}[ht]
  \begin{center}
    \label{tab:comparison}
    \begin{tabular}{|c|c|r|r|} 
    \hline
    Dataset & \textbf{Precision} & \textbf{Recall} & \textbf{F1-score}\\
      \hline

    ORIG-HNFC & 62.7 & 72.1 & 67.1\\
      \hline
    DE-ID-HNFC & 61.7 & 70.2 & 65.7\\
    \hline
    \end{tabular}
  \end{center}
  \caption{ICD-10 association results with~\cite{dalloux2020supervised} approach before and after de-identification}
  
  \label{result}
\end{table}

{The results presented in the table show that our method of identification degrades the results (precision and recall) in a very slight degree (less than 2.6\%).  
Given the very low difference, it means that our de-identification method preserves the structure and consistence of the documents. Although this difference is not great, it confirms the importance of the document's readability, the coherence of the document and the necessity to keep a maximum of information (date, age, location...) for a fine analysis.}




\section{Conclusion and Future Work}\label{section:Conclusion}
This work presented a global method for de-identifying textual medical documents adapted to the French language.
The HIPAA-defined entity recognition (NER) of medical documents is based on a combination of rule-based approaches implemented in MEDINA and deep learning-based approaches using FlauBERT transformer and NERDA.
The substitution of these entities with plausible privacy features is either random when it has no impact on the document, or based on local differential privacy (LDP), a theoretical framework accepted as a de facto standard in privacy.

The robustness of the approach was evaluated on an original medical dataset within a French public hospital.  
In terms of entity detection, the selected combination is the most efficient to date.
In terms of privacy protection, the selected $\epsilon$-LDP mathematically guarantees that the addition of noise is largely sufficient to make re-identification impossible, if not very difficult. This is the first time that such a property has been established on medical documents.

The utility of the approach was globally observed on a dataset by exhaustively comparing the ICD-10 codes associated with and without this de-identifica\-tion step. The association of these codes is of high quality, more accurate, but not systematically identical with and without de-identification, the treatment of dates being perhaps very/too protective for the patient.

Let us continue with future work.
On manually annotated documents in French to be built, we first think of implementing machine learning to distinguish dates ("15 years ago") from ages ("she was 15 years old"), to allow a more accurate de-identification than the one currently implemented.

Regarding NES, we are thinking of implementing a metric-based LDP algorithm to substitute one location for another with similar epidemiological properties. Similarly, this context could also be applied to dates and avoid the medically sensible but non-linear division of short term, medium term, etc.

\bibliographystyle{plain}

\bibliography{references}
\end{document}